\documentstyle[12pt,psfig,epsf]{article}

\begin{document}

\baselineskip=20pt
  
\begin{titlepage}
\begin{flushright}
LBL-38362
\end{flushright}
\vspace{0.6in}

\begin{center}
{\Large\bf
Pion electromagnetic form factor\\ at finite temperature}\\
\vspace{0.2in}
Chungsik Song and Volker Koch\\
\vspace{0.2in}
{\small\it Nuclear Science Division, MS 70A-3307\\ 
Lawrence Berkeley National Laboratory, Berkeley, CA 94720, USA }\\
\vspace{0.1in}
\today \\
\vspace{0.3in}

{\bf Abstract}
\end{center}
Temperature effects on the electromagnetic couplings of
pions in hot hadronic matter are studied with an effective chiral Lagrangian. 
We show that the Ward-Takahashi identity is satisfied at non-zero temperature 
in the soft pion limit.
The in-medium electromagnetic 
form factor of the pion is obtained in the time-like region and 
shown to be reduced
in  magnitude, especially near the vector-meson resonance region. 
Finally, we discuss the consequences of this medium effect on dilepton 
production from hot hadronic matter. 
\vspace{.3in}

\noindent
PACS: 25.75.+r, 14.40.Aq, 12.39.Fe, 12.40.Vv
\end{titlepage}

\section{Introduction}

It is anticipated that there will be 
a phase transition in quantum chromodynamics (QCD) 
at very high temperatures.
At high temperatures a hadronic system  
would be in a plasma phase consisting of weakly interacting 
quarks and gluons, the quark-gluon plasma (QGP),
while at low temperatures hadronic matter is described well by 
mesons and baryons. Chiral symmetry, a symmetry of QCD in the limit of 
massless quarks, is spontaneously broken in the ground state of QCD 
as evidenced in the
small mass of the pion. At high temperatures, above the phase transition,
chiral symmetry is expected to be restored \cite{qgp}, 
as demonstrated by lattice gauge calculations. 
The formation and observation of this new phase of hadronic 
matter is the  main goal of experiments with high energy
nucleus-nucleus collisions \cite{rhic}. 

Photon and lepton pair production have been suggested as promising probes 
to study the properties of hot hadronic matter \cite{dilepton}.
The strong temperature dependence of 
the production rate of these signals 
makes it possible to discriminate the various states of hadronic matter with
different temperatures. 
Furthermore, they can carry information of the hot matter 
without further distortion since these electromagnetic probes interact very
weakly with surrounding particles.

Dilepton production from the low temperature hadronic phase has been 
considered
a possible probe for the chiral phase transition.
In hot hadronic matter, even below the phase transition temperature, chiral
symmetry is expected to be partially restored, i.e. the magnitude of the order
parameter $<\bar{q} q>$ is reduced from its vacuum value.
As a result the properties of light mesons, in particular the vector mesons,
might be modified, and these changes will affect the dilepton spectrum. 
It has been suggested that the masses of vector mesons would change as the 
hadronic matter undergoes a phase transition to the chirally symmetric phase
\cite{pisaski,br}. 
If this is the case, one should be able to observe this effect directly
through the shift of vector meson peaks in the dilepton spectra from
heavy ion collisions.

A recent study based on partial conservation of axial-vector current (PCAC) 
and current algebra, however, has shown
that up to $T^2$, where $T$ is the temperature,
there is no change in vector meson masses but
only a mixing between the vector and axial vector correlators \cite{dei}.
This result should  
be satisfied by any models that include 
the symmetry properties of low energy hadronic physics. 
Indeed, results from effective chiral Lagrangian approaches 
\cite{slee,song1,pisaski2}
are consistent with this temperature dependence, and 
vector meson masses obtained from these models do not change appreciably 
unless the temperature of the hadronic matter is very 
close to the critical temperature for the phase transition. 

Instead, the mixing of the vector and axial-vector correlation function at
finite temperature implies that the leading effect of the temperature in vector
meson properties appears 
in the coupling of vector mesons to photon, which goes like $T^2$.
Recently it has been suggested
this effect will affect the dilepton yield from hot hadronic matter
produced in high energy nucleus-nucleus collisions \cite{song2}.
In hot hadronic matter, the production of dileptons with invariant masses 
near the $\rho$ resonance is dominated by 
pion-pion annihilation.  According to the vector meson dominance 
(VMD) assumption \cite{sak}, two pions in this process
form a rho meson that subsequently converts into a virtual photon. 
The dilepton yield depends, thus, on the 
pion electromagnetic form factor,
\begin{equation}
F_\pi(q^2)={{g_{\rho\pi\pi} g_{\rho\gamma}}
\over m_\rho^2-q^2-im_\rho\Gamma_\rho},
\end{equation}
where $g_{\rho\gamma}$ is the photon-$\rho$-meson coupling constant, 
$g_{\rho\pi\pi}$ is the pion-$\rho$-meson coupling constant, 
and $\Gamma_\rho$ is the neutral $\rho$ meson decay width. 
This form factor has been extensively used in calculating
the dilepton emission rate from hadronic matter at finite temperature.
In these studies, the form factor has been taken to be independent
of temperature. 
However, the change of the photon-vector meson coupling in medium indicates
that the $F_\pi(q^2)$ is to be modified at finite temperature,
and this will affect on dilepton production 
in hot hadronic matter.
 
In the present paper, we shall study the pion electromagnetic form 
factor at finite temperature using an effective chiral Lagrangian
that includes explicitly the vector mesons and gives also the correct 
mixing effect at finite temperature. In section 2 we include details
for the Lagrangian we use in the paper.
In section 3 we summarize the modification of pion-photon couplings in hot
hadronic matter.
We find that the pion-photon coupling is affected in medium not only
because the photon-$\rho$-meson 
coupling is modified at finite temperature
but also because the $\pi - \rho$ coupling as well as the properties of the
$\rho$-meson itself are changed in  hot hadronic matter. 
In section 4 we study the changes of electromagnetic couplings of pions at
low temperatures in the soft pion limit, where pion and photon momenta are
assumed to be small compared to vector meson mass.
We also calculate the 
in-medium pion electromagnetic form factor near vector resonance
region. Here we consider the corrections to pion-photon coupling in hot
hadronic matter with one loop diagrams. The one loop corrections are regarded
as the dominant one since the 
density of most hadrons is small at the temperature we are interested in.
In section 5 we study the 
effect on the dilepton production rate from $\pi-\pi$ annihilation in 
hot matter, using the temperature-dependent form factor.

\section{Hadrons at low energy $-$ Effective field theory}

There is a general agreement at the present time that quantum chromodynamics
(QCD) is the correct theory of strong interactions. Although QCD is simple and
elegant in its formulation, the derivation of its physical predictions for low
energy phenomena present,
however, arduous difficulties because of long distance QCD effects that are
essentially non-perturbative. The theoretical progress has gone through
various directions, including lattice simulations, the use of sum rules, and
employing effective field theory.

In the effective field theory we regard mesons and baryons as 
elementary particles and construct a Lagrangian
with the symmetry of the fundamental theory.
In QCD chiral $SU(N_f)\times SU(N_f)$ symmetry, current algebra and PCAC
play an important role to construct an effective Lagrangian. 
This effective theory can describe all of the couplings in terms of a
relatively small number of parameters 
and is very successful for low energy hadron physics. For example,     
chiral perturbation theory provides a compact and elegant
method for dealing with the interactions of pions at low energies. 
This approach is reliable, however, only when the internal structure of
hadrons, i.e., quark and gluon content of hadrons, can be neglected. 


Another important aspect of hadron physics for 
the present work is that the interaction
of hadrons to the photon. This has been remarkably well described
by using the vector meson dominance assumption. 
This assumes that the hadronic
components of the vacuum polarization of the photon consist exclusively of the
known vector mesons.  At energies below 1GeV the neutral vector mesons,
$\rho^0,\omega$ and $\phi$ play an important role in the electromagnetic
interactions of hadrons. The pion electromagnetic form factor is a particularly
striking example of the VMD model.   
The concept of VMD, of course, is purely phenomenological and has not yet been
proven from the fundamental theory. 


Hidden local symmetry (HLS) is a natural framework for 
describing the vector mesons
in a manner consistent with chiral symmetry of QCD and vector meson dominance
assumption \cite{hls}. 
The HLS Lagrangian yields
at tree level a successful phenomenology for the pions and $\rho$ mesons. Let
us start with the $G_{global}\times H_{local}$ ``linear model" with
$G=SU(2)_L\times SU(2)_R$ and $H=SU(2)_V$. 
It is constructed with two 
SU(2)-matrix valued variables $\xi_L(x)$ and $\xi_R(x)$, which transform 
as $\xi_{L,R}(x) \rightarrow \xi'_{L,R}(x)=h(x) \xi_{L,R}~ g^\dagger_{L,R}$ 
under
$h(x)\in[$SU(2)$_V]_{\rm local}$ and $g_{L,R}\in[$SU(2)$_{L,R}]_{\rm global}$.

Introducing the vector meson $V_\mu$  as the gauge field of the local 
symmetry and the photon ${\cal B}_\mu$ as an external gauge field of the global
symmetry, we have the following chirally invariant Lagrangian,
\begin{eqnarray} 
{\cal L}& = & -{f_\pi^2\over4} {\rm tr} \left[{\cal D}_\mu \xi_L \cdot 
         \xi_L^\dagger - {\cal D}_\mu \xi_R \cdot \xi_R^\dagger \right]^2 
 \nonumber \\ [12pt] 
      & & - {af_\pi^2\over4} {\rm tr} \left[{\cal D}_\mu \xi_L 
      \cdot \xi_L^\dagger + {\cal D}_\mu \xi_R \cdot \xi_R^\dagger  \right]^2
      + {\cal L}_{\rm kin} ( V_\mu ,{\cal B}_\mu),
\label{l_hidden}
\end{eqnarray}
where $a$ is an arbitrary constant and 
$f_\pi=93\>\rm MeV$ is the pion decay constant.
The covariant derivative is given by 
\begin{eqnarray} 
{\cal D}_\mu \xi_{L,R}= (\partial_\mu -igV_\mu)\xi_{L,R}
                        + ie \xi_{L,R} {\cal B}_\mu\tau_3/2,
\end{eqnarray}
where $\tau_3$ is the isospin Pauli-matrix.  
The kinetic terms, ${\cal L}_{\rm kin} ( V_\mu ,{\cal B}_\mu)$, are
conventional non-Abelian and Abelian gauge field tensors for vector meson 
and photon field, respectively. 
In order to obtain masses for the pseudoscalar mesons, 
we introduce an explicit symmetry
breaking term, ${\cal L}_{\rm SB} (\xi_{L,R})$, which is given by
\begin{equation}
{\cal L}_{\rm SB}(\xi_{L,R})={1\over4}f_\pi^2m_\pi^2
{\rm tr}(\xi_L\xi_R^\dagger+\xi_R\xi_L^\dagger). 
\end{equation}

\begin{figure}
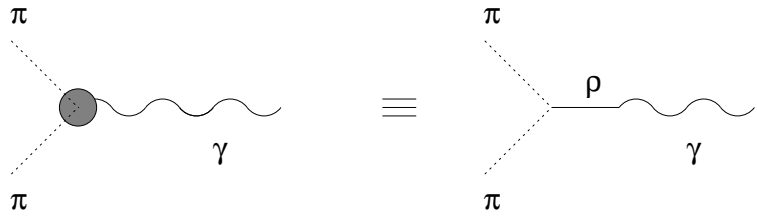

\caption{Vector meson dominance in pion electromagnetic form factor}
\label{vmd}
\end{figure}

In the ``unitary" gauge, 
\begin{eqnarray}
\xi_L^\dagger(x)=\xi_R(x)=e^{i\pi(x)/f_\pi} \equiv \xi(x),
\end{eqnarray}
the effective Lagrangian takes the form,
\begin{eqnarray}
{\cal L} & = & -\frac{1}{4}(F^{V}_{\mu \nu})^2 -\frac{1}{4} 
     (\partial_\mu {\cal B}_\nu- \partial_\nu {\cal B}_\mu)^2
     +\frac{1}{4} {\rm tr}(\partial_\mu U \partial^\mu U^\dagger)
     +\frac{1}{2} m_\rho^2 V_\mu^2-eg_\rho V_3^\mu {\cal B}_\mu
\\[12pt]\nonumber
   & & + g_{\rho  \pi \pi} V^\mu \cdot (\pi \times \partial_\mu \pi)
       + g_{\gamma\pi \pi} {\cal B}^\mu  (\pi \times \partial_\mu \pi)_3
      +\cdots
\end{eqnarray}
where $U=\xi^2(x)$ and the parameters are given as 
\begin{eqnarray}
m_\rho^2 &=& ag^2f_\pi^2,\cr
g_{\rho\pi\pi} &=& {1\over2}ag,\cr
g_{\rho\gamma} &=& agf_\pi^2,\cr
g_{\gamma\pi\pi} &=& (1-{1\over2}a)e.
\end{eqnarray}
With $a=2$, we have 
the universality of the $\rho$-couplings ($g_{\rho\pi\pi}=g$), 
the Kawarabayashi-Suzuki-Riazuddin-Fayyazuddin (KSRF) relation
$m_\rho^2 = 2 g_{\rho \pi \pi}^2 f_\pi^2$,
and the $\rho$ meson dominance of the 
pion-photon coupling ($g_{\gamma \pi \pi}=0$).   
In this effective Lagrangian, the pion electromagnetic form factor in free
space can be obtained at tree level from the diagram shown in Fig.~\ref{vmd}.
One sees that the vector meson dominance appears
naturally. A photon converts into a rho meson which interacts
with the pion. The resulting pion electromagnetic form factor is exactly the
same form as (1) assumed in VMD. 

\section{Electromagnetic coupling of pions in the medium}

At non-zero temperature the couplings of pions to the electromagnetic field is
modified by the interaction with particles in the heat bath.
These effects can be included by  thermal loop corrections to the vertices.
In a low temperature pion gas it is possible to expand the medium effect 
as a series of the power in $T^2/f_\pi^2$.
However, we cannot apply the same approximation at the temperatures
100 MeV$< T\le T_c$. 
Instead, we use the fact that the density of particles 
are small even at temperatures near $T_c$. 
In this case we can expand the thermal corrections 
by the number of loops and include one-loop diagrams as leading terms.
We neglect the contributions of vector meson loops in the present calculation, 
since they are suppressed by Boltzmann factors $\sim e^{-m_V/T}$ with 
large masses $m_V \gg T$.

\subsection{Vector and Axial-Vector Mixing}

It has been shown that chiral symmetry predicts
a mixing between vector and
axial vector current-current correlators at low temperature \cite{dei}. 
This mixing implies that the difference
of vector and axial vector current correlator vanishes 
with increasing temperature, which is a 
consequence of the chiral symmetry restoration in 
hadronic matter at finite temperature \cite{ks}.

\begin{figure}
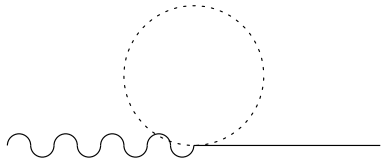

\caption{Correction to the photon-vector meson coupling at finite temperature.
         Here and in the following the dotted, solid and wave lines indicate
         the pions, vector mesons and electromagnetic fields, respectively.}
\label{mixing}
\end{figure}

The mixing effect in vector and axial-vector correlator at finite temperature
has also been studied in the effective Lagrangian approach \cite{slee,song1}.
The interaction with thermal pions as shown in fig. \ref{mixing} is responsible
for the mixing effect at finite temperature. 
The effect of the mixing on the pion electromagnetic 
form factor results in a change of
the photon-vector meson coupling, $g_{\gamma\rho}(T)$, at finite temperature;
\begin{equation}
g_{\gamma\rho}(T)=(1+\epsilon)g_{\gamma\rho}(0),
\label{grho}
\end{equation}
where 
\begin{eqnarray}
\epsilon &=& {1\over f_\pi^2}T\sum_{n_l}\int {d^3l\over (2\pi)^3}
                                       {1\over l^2-m_\pi^2}\cr
         &=& -{T^2\over 12f_\pi^2} \qquad\mbox{(in chiral limit)}.
\end{eqnarray}
This implies that the vector-photon coupling will be reduced in hot matter due
to the mixing of vector and axial vector currents.  Here the sum is over 
the Matsubara frequency of thermal pions which is given by
$\omega_l=2\pi Tn_l$ with integer $n_l$. 

\subsection{Vector meson properties}

\begin{figure}
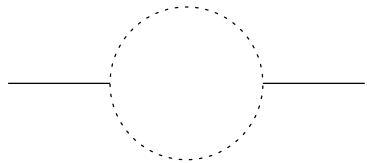

\caption{Vector meson propagator at finite temperature}
\label{rho}
\end{figure}

With the vector meson dominance assumption the pion couples to the 
electromagnetic 
field only through a rho meson intermediate state . 
Thus the changes in the rho meson
properties in medium also  affect  the pion-photon coupling. 
To include the in medium rho properties we
calculate the vector meson self-energy with an one-loop diagram in
fig.~\ref{rho}.
The explicit form is given by 
\begin{equation}
\Pi_\rho^{\mu\nu}=-({1\over2}ag)^2T\sum_{n_l}\int{d^3l\over(2\pi)^3}
{(2l^\mu-k^\mu)(2l^\nu-k^\nu)\over(l^2-m_\pi^2)((l-k)^2-m_\pi^2)},
\label{rho_self}
\end{equation}
to leading order, 
where $k$ is the momentum of vector meson.
The in-medium propagator is then given by 
\begin{eqnarray}
D_{\mu\nu} &=& D^0_{\mu\nu}
              +D^0_{\mu\lambda}(-i\Pi^{\lambda\sigma}) D_{\sigma\nu}\cr
           &=& D^0_{\mu\nu}
              +D^0_{\mu\lambda}(-i\Pi^{\lambda\sigma}) D^0_{\sigma\nu}\ldots
\end{eqnarray}
where $D^0_{\mu\nu}$ is the propagator in free space. 

The modification of pion-photon vertex due to the change of the vector meson
propagator in medium can be written as 
\begin{eqnarray}
\Gamma^{rho}_\mu &=& -ig_{\rho\gamma}
D^0_{\mu\lambda}(-i\Pi^{\lambda\sigma}) D^0_{\sigma\nu}g_{\rho\pi\pi}
(p_\nu-q_\nu)+\cdots
\nonumber \\ [12pt] 
&=&-{1\over4}a^2
(a g^2 f_\pi^2)^2{g_{\mu\nu}-k_\mu k_\nu/m_\rho^2\over k^2-m_\rho^2} 
 \nonumber \\ [12pt] 
&&\times
{1\over f_\pi^2}T\sum_{n_l}\int{d^3l\over(2\pi)^3}
(2l^\nu-k^\nu){l\cdot(p-q)\over (l^2-m_\pi^2)((l-k)^2-m_\pi^2)}
\times{1\over k^2-m_\rho^2}+\cdots,
\end{eqnarray}
where $p$ and $q$ are momenta of external pions and $k=p+q$. 

\subsection{Vertex corrections}

\begin{figure}
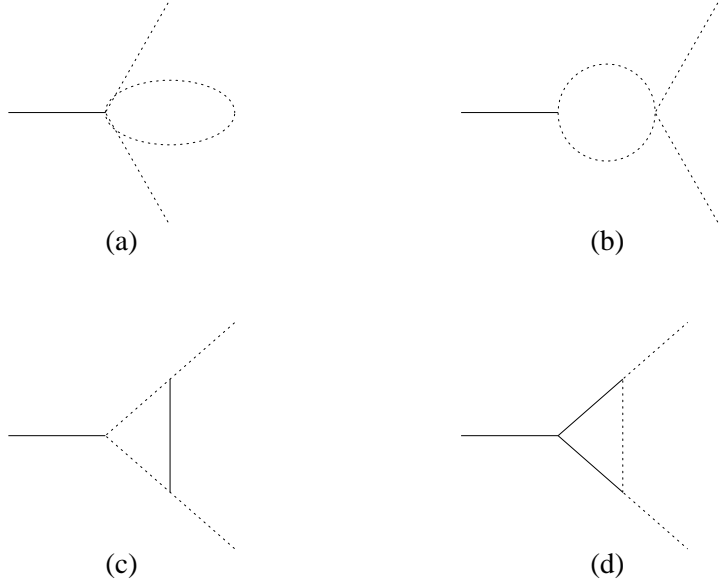

\caption{Correction to the $\pi-\pi-\rho$ vertex at finite temperature}
\label{vertex}
\end{figure}

We consider the thermal effect on the $\pi-\pi-\rho$ coupling in hot hadronic
matter which is given by fig.~\ref{vertex}. 
Each contribution to the pion-photon coupling is given by
\begin{eqnarray}
\Gamma^{(4-a)}_\mu &=& -(p_\mu-q_\mu)
{a g^2 f_\pi^2\over k^2-m_\rho^2} {5a\over24}
{1\over f_\pi^2}T\sum_{n_l}\int{d^3l\over(2\pi)^3}{1\over l^2-m_\pi^2},
 \nonumber \\ [12pt] 
\Gamma^{(4-b)}_\mu &=& -{1\over2}a\left({3a\over4}-1\right)
a g^2 f_\pi^2{g_{\mu\nu}-k_\mu k_\nu/m_\rho^2\over k^2-m_\rho^2} 
 \nonumber \\ [12pt] 
&&\times
{1\over f_\pi^2}T\sum_{n_l}\int{d^3l\over(2\pi)^3}
(2l^\nu-k^\nu){l\cdot(p-q)\over (l^2-m_\pi^2)((l-k)^2-m_\pi^2)},
 \nonumber \\ [12pt] 
\Gamma^{(4-c)}_\mu &=& -{1\over2}({1\over2}a)^2
(a g^2 f_\pi^2)^2{g_{\mu\nu}-k_\mu k_\nu/m_\rho^2\over k^2-m_\rho^2} 
 \nonumber \\ [12pt] 
&&\times 
{1\over 2f_\pi^2}T\sum_{n_l}\int{d^3l\over(2\pi)^3}\Biggl[
(2l^\nu-k^\nu)
{(l+p)_\lambda
(g^{\lambda\sigma} -(l^\lambda-p^\lambda)(l^\sigma-p^\sigma)/m_\rho^2)
(l_\sigma-q_\sigma-k_\sigma)
\over(l^2-m_\pi^2)((l-k)^2-m_\pi^2)((l-p)^2-m_\rho^2)}
 \nonumber \\ [12pt] 
&&\qquad\qquad\qquad -(2l^\nu-k^\nu)
{(l_\lambda+q_\lambda)
(g^{\lambda\sigma} -(l^\lambda-q^\lambda)(l^\sigma-q^\sigma)/m_\rho^2)
 (l_\sigma-p_\sigma-k_\sigma)
\over(l^2-m_\pi^2)((l-k)^2-m_\pi^2)((l-q)^2-m_\rho^2)}
\Biggr],
 \nonumber \\ [12pt] 
\Gamma^{(4-d)}_\mu &=& {1\over2}a
(a g^2 f_\pi^2)^2{g_{\mu\nu}-k_\mu k_\nu/m_\rho^2\over k^2-m_\rho^2} 
 \nonumber \\ [12pt] 
&&\times 
{1\over f_\pi^2}T\sum_{n_l}\int{d^3l\over(2\pi)^3}\Biggl\{
\biggl[ (2l^\nu+p^\nu-q^\nu)g^{\alpha\beta}
 -(l^\alpha+k^\alpha+q^\alpha)g^{\beta\nu}
 -(l^\beta-p^\beta-k^\beta)g^{\nu\alpha}\biggr]
 \nonumber \\ [12pt] 
&&\qquad\qquad\qquad\times
(g_{\alpha\alpha'}-(l_\alpha-p_\alpha)(l_{\alpha'}-p_{\alpha'})/m_\rho^2)
(g_{\beta\beta'}-(l_\beta+q_\beta)(l_{\beta'}+q_{\beta'})/m_\rho^2)
 \nonumber \\ [12pt] 
&&\qquad\qquad\qquad\times
{(l^{\alpha'}+p^{\alpha'})(l^{\beta'}-q^{\beta'})
  \over(l^2-m_\pi^2)((l-p)^2-m_\rho^2)((l+q)^2-m_\rho^2)}\Biggr\}.
\end{eqnarray}

\subsection{Direct coupling}

\begin{figure}
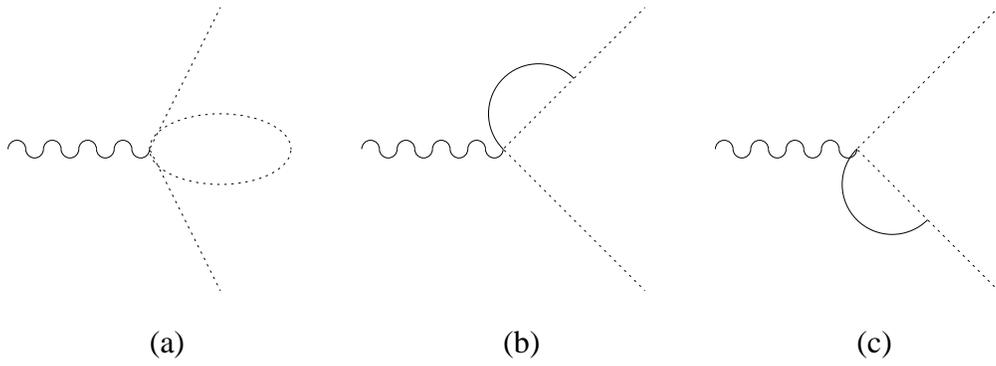

\caption{The $\pi-\pi-\gamma$ vertex at finite temperature}
\label{direct}
\end{figure}

In medium it is possible for pions to couple to the photon fields directly
as shown in fig.~\ref{direct}.
The interaction with thermal pions make it possible for pions to 
couple to the photon, which is forbidden in free space. 
This implies that strict vector meson dominance is not satisfied at
non-zero temperature. 
Each contribution is given by 
\begin{eqnarray}
\Gamma^{(5-a)}_\mu &=& (p_\mu-q_\mu)
{5\over3}\left(1-{7a\over8}\right)
{1\over f_\pi^2}T\sum_{n_l}\int{d^3l\over(2\pi)^3}{1\over l^2-m_\pi^2},
 \nonumber \\ [12pt] 
\Gamma^{(5-b)}_\mu + \Gamma^{(5-c)}_\mu 
&=& -{3a\over4}a g^2 f_\pi^2
{1\over f_\pi^2}T\sum_{n_l}\int{d^3l\over(2\pi)^3}
 \nonumber \\ [12pt] 
&&\times
\Biggl[ {(l^\mu+p^\mu)\over (l^2-m_\pi^2)((l-p)^2-m_\rho^2)}
      -{(l^\mu+q^\mu)\over (l^2-m_\pi^2)((l-q)^2-m_\rho^2)}\Biggr].
\end{eqnarray}

\subsection{Pion wave function renormalization}
 
Finally, there is a contribution from the modification of pion properties
in hot matter. 
As pions propagate through the medium they couple to particles in 
the thermal bath and their properties will be modified. 
The medium effect on the pion propagator can be included in the self-energy
which is defined as the difference 
between the inverse of the in-medium propagator ($D^{-1}$) 
and that of the vacuum propagator $(D_0^{-1})$:
\begin{equation}
-i\Pi=D^{-1}-D_0^{-1}.
\end{equation}

Such a modification renormalizes the pion wave function and affects  the
strength of $\pi-\pi$ annihilation into a  vector meson.
At zero temperature the wave function renormalization constant can be uniquely
defined. At finite temperature, however, it cannot, because 
Lorentz invariance is broken due to
the presence of the heat bath. Thus the self-energy may have separate
dependences on the momentum and energy. 
Here, the wave function renormalization constant will be defined as 
\begin{equation}
Z_\pi^{-1}=1-{\partial\Pi_\pi\over\partial p_0^2}
\Biggl\vert_{\vec p=0,p_0=m_\pi}.
\label{zpi}
\end{equation}

\begin{figure}
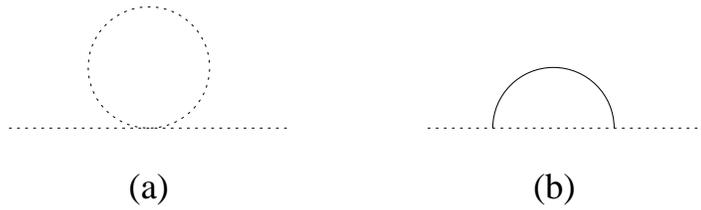

\caption{Pion self-energy at finite temperature}
\label{pion}
\end{figure}

We calculate the pion self-energy from the one loop diagrams shown in
fig.~\ref{pion}. 
Each contribution is given by 
\begin{eqnarray}
\Pi_\pi^{(6-a)}(p) &=& {1\over 6f_\pi^2}T\sum_{n_l}\int{d^3l\over(2\pi)^3}
\left[5m_\pi^2-4\left(1-{3\over4}a\right)(p^2+l^2)\right]{1\over l^2-m_\pi^2},
\nonumber \\ [12pt] 
\Pi_\pi^{(6-b)}(p) &=& {1\over2}a(ag^2f_\pi^2){1\over f_\pi^2}
T\sum_{n_l}\int{d^3l\over(2\pi)^3}
{(l+p)^2+(l^2-p^2)^2/m_\rho^2\over (l^2-m_\pi^2)((l-p)^2-m_\rho^2)}.
\label{pion_self}
\end{eqnarray}

\section{Pion electromagnetic form factor at $T\ne 0$}

\subsection{Effective charge of pions}

First let us calculate the effective charge of a pion in medium by
considering scattering of a photon off the pion.
We consider the soft pion limit in which four momenta of  
pions are assumed to be small compared to the vector meson mass. 
Actually we approximate that 
\begin{equation}
(l-p)^2,\quad (l-q)^2 \ll m_\rho^2
\end{equation}
where $l$ is the momentum of  pion in a thermal loop and $p$ and $q$ 
are external pion momenta.
Since the momenta in a thermal loop are of the order of the
temperature, this approximation should be reasonable for low temperatures and
small external pion momenta.
We include only thermal pion effects and expand the medium effect as a power
series of $T^2/f_\pi^2$. 

With this soft pion approximation, we obtain very simple expressions for vertex
corrections. For example, we have 
\begin{eqnarray}
\Gamma^{(4-c)}_\mu &\approx&  -{1\over2}({1\over2}a)^2
(a g^2 f_\pi^2)^2{g_{\mu\nu}-k'_\mu k'_\nu/m_\rho^2\over k'^2-m_\rho^2} 
 \nonumber \\ [12pt] 
&&\times 
{1\over (-m_\rho^2)}{1\over 2f_\pi^2}T\sum_{n_l}\int{d^3l\over(2\pi)^3}\Biggl[
(2l^\nu+k'^\nu)
{2l\cdot(p'+p)+k'\cdot(p'+p)\over(l^2-m_\pi^2)((l+k')^2-m_\pi^2)}
\Biggr]
\end{eqnarray}
where we use $p'=-q$ and $k'=(p'-p)=-(p+q)$.
The vertex correction is then given by
\begin{eqnarray}
\Gamma_\mu^{vertex} &=& -(p'_\mu+p_\mu)
{5a\over 24f_\pi^2}{ag^2f_\pi^2\over k'^2-m_\rho^2}
                 T\sum_{n_l}\int{d^3l\over(2\pi)^3}{1\over l^2-m_\pi^2}
\nonumber \\ [12pt] 
                   &&+{1\over2}a
(a g^2 f_\pi^2){g_{\mu\nu}-k'_\mu k'_\nu/m_\rho^2\over k'^2-m_\rho^2} 
 {1\over f_\pi^2}
               T\sum_{n_l}\int{d^3l\over(2\pi)^3}
 {(2l^\nu+k'^\nu)\over
               (l^2-m_\pi^2)((l+k')^2-m_\pi^2)}
\nonumber \\ [12pt] 
&&\qquad\qquad\times
              \left[ \left(1-{1\over2}a\right)l\cdot(p'+p)
                    +{1\over8}ak'\cdot(p'+p)\right]
\Biggr\}. 
\label{rpp_soft}
\end{eqnarray}
where the diagram (4-d) is not included since the contribution will be 
suppressed by the factor of ${p^2/m_\rho^2}$.
With $a=2$ the contributions  $\Gamma_\mu^{(4-b)}$ and
$\Gamma_\mu^{(4-c)}$ cancel each other.  
With the same approximation we also have a simple form 
for the direct coupling in medium: 
\begin{equation}
\Gamma^{direct}_\mu = (p'_\mu+p_\mu)
                 \left({5\over3}-{17a\over24}\right)
                 {1\over f_\pi^2}
                 T\sum_{n_l}\int{d^3l\over(2\pi)^3}{1\over l^2-m_\pi^2}.
\end{equation}
The total modification of the  coupling of pions to the photon field 
in medium now can be
written as  
\begin{eqnarray}
\Gamma^{\gamma\pi\pi}_\mu(T) &=& 
\Gamma_\mu^{mixing}+\Gamma_\mu^{vertex}+\Gamma_\mu^{direct}+\Gamma_\mu^{rho}
\nonumber \\ [12pt] 
&=&  (p'_\mu+p_\mu) \left({1\over 4f_\pi^2}-
{a g^2 f_\pi^2\over k'^2-m_\rho^2}{17\over 12f_\pi^2}\right)
                 T\sum_{n_l}\int{d^3l\over(2\pi)^3}{1\over l^2-m_\pi^2},
\nonumber \\ [12pt] 
&&+\left({a g^2 f_\pi^2\over k'^2-m_\rho^2}\right){1\over 4f_\pi^2}
               T\sum_{n_l}\int{d^3l\over(2\pi)^3}
               {(2l^\mu+k'^\mu)\>k'\cdot(p'+p)\over
               (l^2-m_\pi^2)((l+k')^2-m_\pi^2)}
\nonumber \\ [12pt] 
&&-\left({a g^2 f_\pi^2\over k'^2-m_\rho^2}\right)^2 
{1\over f_\pi^2}T\sum_{n_l}\int{d^3l\over(2\pi)^3}
\left[{(2l_\mu+k'_\mu)\>l\cdot(p'+p)\over (l^2-m_\pi^2)((l+k')^2-m_\pi^2)}
-{k'_\mu\over m_\rho^2}{k'\cdot(p'+p)\over (l^2-m_\pi^2)}\right]
\nonumber \\ [12pt] 
\label{vertex_soft}
\end{eqnarray}
with $a=2$.

In general the change in vertex function is related to the modification 
of pion propagator by the Ward-Takahashi (WT) identity \cite{drell};
\begin{equation}
(p'_\mu-p_\mu)\Gamma_{\gamma\pi\pi}^\mu(p',p)=\Pi_\pi(p)-\Pi_\pi(p'),
\label{wt}
\end{equation}
where $\Pi_\pi(p)$ is the pion self-energy. 
This is a consequence of the $U(1)$ gauge symmetry of the theory. 
At finite temperature we have from (\ref{vertex_soft})
\begin{equation}
(p'_\mu-p_\mu)\Gamma_{\gamma\pi\pi}^\mu(p',p;T)
={2\over 3f_\pi^2}(p'^2-p^2)
T\sum_{n_l}\int{d^3l\over(2\pi)^3}{1\over l^2-m_\pi^2}
\label{vertex2}
\end{equation}
in the limit $k'^2\to 0$.
On the mass shell this just shows that the electromagnetic 
current is conserved at finite temperature. 
The pion self-energy can also be calculated in the soft pion limit from
(\ref{pion_self}) and it is given by
\begin{equation}
\Pi_\pi={1\over 6f_\pi^2}T\sum_{n_l}\int{d^3l\over(2\pi)^3}
{1\over l^2-m_\pi^2}( 5m_\pi^2-4l^2-4p^2 ).
\label{self-energy}
\end{equation}
From these two results, i e., (\ref{vertex2}) and (\ref{self-energy}),
one can easily see that Ward-Takahashi identity is 
satisfied at finite temperature. 

At zero temperature the WT identity 
implies that the vertex correction in charge renormalization is exactly
canceled by the wave function renormalization constant.
This is not obvious at finite temperature because of 
the broken Lorentz invariance in the presence of the heat bath.
We must, therefore, be careful when taking  limits 
in order to proceed with the
temperature dependent renormalization of the electric charge.
Since at finite temperature a general amplitude $A(k'_0,\vec k')$ may have 
different functional dependences
on $k'_0$ and $\vec k'$, we should distinguish the limit
$\vec k'=0,k'_0\to 0$ from that of $k'_0=0, \vec k'\to 0$. 

First we consider the limit $\vec k'=0,k'_0\to 0$.
From the WT identity we can show the conservation of  
the effective charge defined by 
\begin{equation}
\Lambda_0(k'_0,k')\vert_{\vec k'=0,k'_0\to 0}\equiv -(p'_0+p_0)e^{eff},
\label{eff1}
\end{equation}
where
\begin{equation}
\Lambda_\mu=-e[(p_\mu'+p_\mu)+\Gamma_\mu]Z_\pi.
\label{lambda}
\end{equation}
From (\ref{wt}) we can get 
\begin{eqnarray}
\Gamma_0(k'_0\to 0,\vec k'=0)&=&-\partial\Pi_\pi(p)/\partial p_0\cr
                      &=&2p_0(Z_\pi^{-1}-1),
\label{gamma0}
\end{eqnarray}
where we use the definition for wave function renormalization constant
(\ref{zpi}) to get the last line.
By inserting (\ref{gamma0}) into (\ref{lambda}) we can see that the effective
charge at finite temperature defined in (\ref{eff1}) 
is conserved as consequence of the WT identity. 

We can show the charge conservation explicitly from the expressions for
the $\pi-\pi-\gamma$ vertex corrections in medium. 
The zeroth component of vertex function to leading order 
can be written as 
\begin{equation}
\Gamma_0^{\rm matt}(p',p)= (p_0'+p_0)
\left[-{5T^2\over 3f_\pi^2}g_0(m_\pi^2/T^2)
+{T^2\over f_\pi^2}g_0(m_\pi^2/T^2)\right],
\end{equation}
where
\begin{equation}
g_0(x)={1\over 2\pi^2}\int_0^\infty{y^2dy\over \sqrt{y^2+x^2}
(e^{\sqrt{y^2+x^2}}-1)}.
\end{equation}
The pion wave function renormalization is obtained from
eq.~(\ref{self-energy}) as    
\begin{eqnarray}
Z_\pi^{-1} &=& 1+{2\over 3f_\pi^2}T\sum_{n_l}
\int{d^3l\over(2\pi)^3}{1\over l^2-m_\pi^2}
\nonumber \\ [12pt] 
      &=& 1-{2T^2\over3f_\pi^2}g_0(m_\pi^2/T^2).
\label{pion_wave}
\end{eqnarray}
There is an exact cancelation at the leading order of $T^2/f_\pi^2$
between the contributions from vertex correction and wave function
renormalization in the definition of the effective charge. 
No temperature dependent correction contributes 
to the charge renormalization from the vertex function. The photon polarization
function is entirely responsible for the charge renormalization. 

If we take the other limit, i.e., $\vec k'=0$ and $k'_0\to 0$, 
the WT identity implies that $\Gamma_i$ instead of
$\Gamma_0$ is related
to the space derivative of the self-energy;
\begin{equation}
\Gamma_i(k'_0=0,k'_j=0,j\ne i;k'_i\to 0)
=-{\partial\Pi_\pi(p)\over\partial p_i}.
\end{equation}
At zero temperature this leads to the same result as that obtained 
in (\ref{gamma0}).
It is not true in general at finite temperature. 
However, in the approximation considered here, the 
pion self-energy depends only on $p^2$, just as in
free space, and we can show the conservation for the effective charge
defined by
\begin{equation}
\Lambda_i(k'_0,k')\vert_{k'_0=0,k'_j=0,j\ne i;k'_i\to 0}
\equiv -(p'_i+p_i)e^{eff}.
\end{equation}

\subsection{Pion electromagnetic form factor in medium}

Now we consider the pion electromagnetic form factor for the process in which 
two pions annihilate into a virtual photon which sub-sequentially decays 
into lepton pairs.
For simplicity we do the calculation in the rest frame of the virtual photon, 
i.e., in the frame where lepton pairs move back-to-back. 
First we use the results obtained in the soft  pion limit 
for couplings of pions to
photon and vector mesons. Even though
this result is reliable only near threshold for two pion annihilation and at
low temperatures, this
gives a good intuition for the form factor at finite temperature. Moreover, we
can do the loop integration exactly in this limit.

We define the pion electromagnetic form factor in medium as follows
\begin{equation}
\Gamma_\mu(T)=(p_\mu-q_\mu)F_\pi(T).
\end{equation} 
The in-medium form factor can be written as 
\begin{equation}
F_\pi(T)=Z_\pi(T)\left[{g_{\rho\pi\pi}(T) g_{\rho\gamma}(T)\over
          m_\rho^2-k^2-i\Gamma_\rho m_\rho-\Pi_\rho(T)}
         +F_\pi^{direct}(T)\right]
\end{equation}
where  $g_{\rho\gamma}(T)$ and $Z_\pi(T)$ are given by (\ref{grho}) and
(\ref{pion_wave}), respectively.
The $g_{\rho\pi\pi}(T)$ is obtained from (\ref{rpp_soft}) as
\begin{equation}
g_{\rho\pi\pi}(T) = g_{\rho\pi\pi}(0)
                       \left(1-{5T^2\over 12f_\pi^2}g_0(m_\pi^2/T^2)\right).
\end{equation}
The last term appears because of the direct coupling of pions to the photon in
medium and is given by 
\begin{equation}
F_\pi^{direct}=-{T^2\over 4f_\pi^2}g_0(m_\pi^2/T^2)
\end{equation}

\begin{figure}
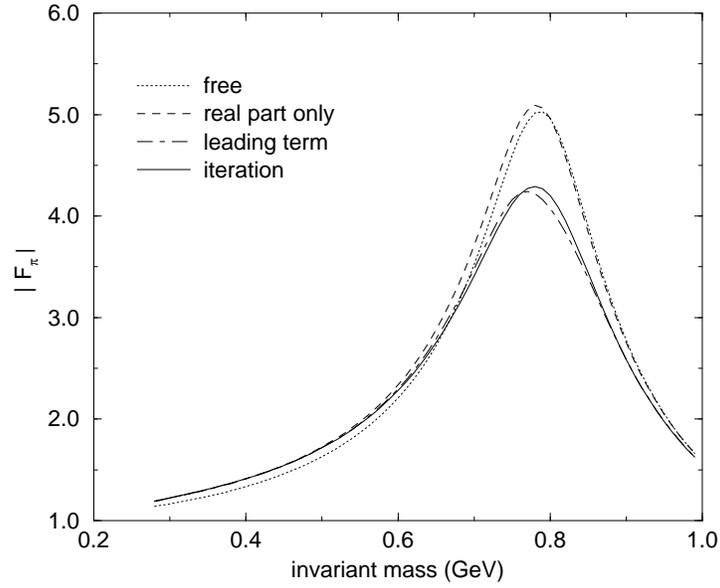

\caption{Pion electromagnetic form factor with the modification in vector
meson properties}
\label{form_vector}
\end{figure}

The $\Pi_\rho(T)$ is 
the modification due to the in-medium rho meson propagator.
In the rest frame of the vector meson  
the propagator is given by
\begin{equation}
D_{\mu\nu} = {g_{\mu\nu}-k_\mu k_\nu\over k^2}
\left({1\over k^2-m_\rho^2+i\Gamma_\rho m_\rho+\Pi_\rho(T)}\right)
+\cdots,
\end{equation}
where $\Pi_\rho(T)$ can be obtained from rho self-energy in 
(\ref{rho_self}) \cite{song4}.
The explicit form of $\Pi(T)$ is given by 
\begin{equation}
\Pi_\rho(T) = {g^2\over 3\pi^2}
       \int{\vert\vec l\vert^2d\vert\vec l\vert\over\omega(e^{\omega/T}-1)}
       {\vert\vec l\vert^2\over\omega^2-k_0^2/4},
\end{equation}
where $\omega=\sqrt{\vert\vec l\vert^2+m_\pi^2}$.
Even though the medium effect on the vector meson mass is small, 
there is an appreciable effect when we include the imaginary part of the rho 
self-energy.
In fig.~\ref{form_vector} we show the effect of vector meson self-energy on the
pion electromagnetic form factor at finite temperature. 
The dashed line is for the result with real part of the
self-energy only and the solid line is that obtained with both real and
imaginary part.
We can see that a large effect comes from the modification of the
imaginary part of the rho self-energy. We  get a comparable  result 
when we restrict ourselves to the leading term in the expansion of 
the self-energy correction (dot-dashed line).
We expect an additional broadening of 
the vector mesons due to collisions \cite{kevin}, which are, however, 
not considered here since they are corrections on the two loop level.

\begin{figure}
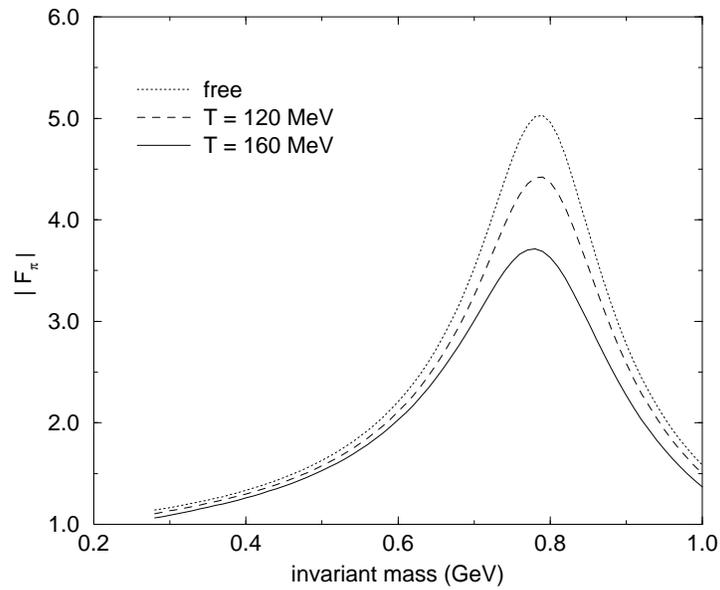

\caption{Pion electromagnetic form factor in soft pion limit}
\label{form_soft}
\end{figure}

The result for pion electromagnetic form factor 
obtained in the soft pion limit 
is shown in fig.~\ref{form_soft} for  different temperatures.
We can see that the form factor is suppressed, particularly near vector 
resonance region, as the temperature increases. 
The reduction of the form factor is mainly due to the suppression
of photon -- $\rho$-meson coupling which comes from the aforementioned  mixing 
effect, and due to the broadening of vector mesons in medium.

Now we consider the pion form factor near the vector resonance region where 
the external pion momentum is not negligible compared to the vector meson mass.
For this case we cannot simply replace $1/((l-p)^2-m_\rho^2)$ in 
the vector propagator by ${1/ m_\rho^2}$. 
When we include the full propagator for the vector meson 
we have 
\begin{eqnarray}
\Gamma^{(4-b)}_\mu &\sim& T\sum_{n_l}\int{d^3l\over(2\pi)^3}
 {l_\mu l\cdot(p-q)\over(l^2-m_\pi^2)((l-k)^2-m_\pi^2)}
\nonumber\\[12pt]
&\equiv& (p_\nu-q_\nu)H^{\mu\nu}(k;T),
\end{eqnarray}
which has been canceled by $\Gamma_\mu^{(4-c)}$ in the soft pion limit.
Since $H_{\mu\nu}$ is a second rank tensor it can be written as
\begin{equation}
H_{\mu\nu}=\alpha A_{\mu\nu}+\beta B_{\mu\nu}
          +\gamma C_{\mu\nu}+\delta D_{\mu\nu}.
\end{equation}
Here $A,B,C$, and $D$ are four independent second rank tensors and are given
as following \cite{kk,heinz}
\begin{eqnarray}
A_{\mu\nu} &=& g_{\mu\nu}-{1\over \vec k^2}
               \left[ k_0( n_\mu k_\nu+n_\nu k_\mu)
                     -k_\mu k_\nu -k^2n_\mu n_\nu\right],\cr
B_{\mu\nu} &=& -{k^2\over \vec k^2}\left(n_\mu-{k_0 k_\mu \over k^2}\right)
                                   \left(n_\nu-{k_0 k_\nu \over k^2}\right),\cr
C_{\mu\nu} &=& -{1\over\sqrt{2}\vert\vec k\vert}
                \left[ \left(n_\mu-{k_0 k_\mu \over k^2}\right)k_\nu
                      +\left(n_\nu-{k_0 k_\nu \over k^2}\right)k_\mu\right],\cr
D_{\mu\nu} &=& {k_\mu k_\nu \over k^2}.
\end{eqnarray}
where $n_\mu$ specifies the rest frame of the hot matter.   
$\alpha, \beta, \gamma$, and $\delta$ are given by 
\begin{eqnarray}
\delta &=& {1\over k^2} k^\mu k^\nu H_{\mu\nu},\cr
\gamma &=& {\sqrt{2}\over\vert\vec k\vert}(k^\mu H_{\mu 0}-k_0\delta),\cr
\beta &=& -{1\over\vec k^2}\left(  k^2 H_{00}
                                  -\sqrt{2}\vert\vec k\vert k_0\gamma
                                  -k_0^2\delta\right),\cr
\alpha &=& {1\over2}(H^\mu_\mu-\beta-\delta).
\end{eqnarray}

In the back-to-back frame where $\vec k=0$ we have 
\begin{eqnarray}
\gamma &=& 0\cr
\alpha &=& \beta.
\end{eqnarray}
The expression for $\Gamma_\mu$ simplifies even more  
when we use on the shell condition for
pions, i.e., $p^2=q^2=m_\pi^2$;
\begin{equation}
\Gamma^{(4-b)}_\mu=(p_\mu-q_\mu)F_\pi(T=0){1\over f_\pi^2}\alpha(T)
\end{equation}
and $\alpha(T)$ is obtained from $H_{\mu\nu}$ as 
\begin{equation}
\alpha(T)= -{1\over 12\pi^2}
        \int{\vert\vec l\vert^2d\vert\vec l\vert\over\omega(e^{\omega/T}-1)}
            {\vert\vec l\vert^2\over\omega^2-k_0^2/4}
\end{equation}

The diagram in fig.~4-c cannot be calculated exactly
with the full propagator of vector mesons. Instead we take reasonable
approximation for the structure of the vertex function $\Gamma_\mu$ and do the
calculation for the loop correction. 
To do this, first, we consider the general form for 
the pion-photon vertex function at finite temperature which is given by 
\begin{equation}
\Gamma_\mu(p,q)=(p_\mu-q_\mu)F+n_\mu G + (p_\mu+q_\mu)G'
\end{equation}
where $F,G$ and $G'$ are scalar functions.
When we use the fact that the 
electromagnetic current should be conserved, that is,
\begin{equation}
(p^\mu+q^\mu)\Gamma_\mu(p,q)=0,
\end{equation}
we can describe the vertex function only 
with two independent functions, $F$ and $G$, as 
\begin{equation}
\Gamma_\mu(p,q)=(p_\mu-q_\mu)F+\left(n_\mu -{k\cdot n\over k^2}k_\mu\right)G 
\end{equation}
where $k=(p+q)$.
In the rest frame of the heat bath, $n_\mu=(1,0,0,0)$, we have 
\begin{eqnarray}
\Gamma_0 &=& (p_0-q_0)F-{\vec k^2\over k^2}G\cr
\Gamma_i &=& (p_i-q_i)F-{k_0 k_i\over k^2}G.
\end{eqnarray}

Now we consider the vertex function in the back-to-back frame where 
$\vec k=0$ and
$p_0=q_0$. In this frame
the vertex
function is given by 
\begin{equation}
\Gamma_\mu(p,q)=(p_\mu-q_\mu)F
\label{eq49}
\end{equation}
as long as the function $G$ has no singularity at the limit $\vec{k} = 0$.
We can show explicitly that the function $G$ is regular as $\vec k\to 0$ in the
soft pion limit. Thus, it is probably safe to assume that the same is true 
when we include the full propagator for the vector mesons.
We therefore write
\begin{equation}
\Gamma^{(4-c)}_\mu\approx (p_\mu-q_\mu)F^{(4-c)}.
\end{equation}
With this approximation we can do the loop calculation and the details 
are given  in appendix A.
The modification of form factor due to the vertex correction is then  
\begin{equation}
F_\pi^{(4-c)}=F_\pi(T=0)
\left\{{g^2\over (p-q)^2}
\biggl[f_1(p,q;T)+(2m_\pi^2+3k^2/2)g_1(p,q;T)+g_2(p,q;T)\biggr]\right\}
\end{equation}
where $f_1$ and the $g_i$'s are given in appendix A.

We should also include the diagram (4-d) which has been neglected due to the
suppression factor $p^2/m_\rho^2$. We can do the integration with the
approximation used for $\Gamma^{(4-c)}$ and details are given in Appendix B. 
Finally, the vertex correction is given by 
\begin{eqnarray}
F_\pi^{vertex} &=& F_\pi(T=0)
\Biggl\{-{5T^2\over 12 f_\pi^2} g_0(m_\pi^2/T^2)+{1\over f_\pi^2}\alpha(T)
\nonumber\\[12pt]
&&\qquad\qquad
      +{g^2\over(p-q)^2}
       \biggl[f_1(p,q;T)+(2m_\pi^2+3k^2/2)g_1(p,q;T)+g_2(p,q;T)\biggr]
\nonumber\\[12pt]
&&\qquad\qquad
      -2g^2\Biggl({1\over(p-q)^2}\biggl[2h_2(p,q;T)-k^2 h_1(p,q;T)\biggr]
\nonumber\\[12pt]
&&\qquad\qquad\qquad\qquad
      +h_1(p,q;T)+(2m_\pi^2-3k^2/2)h_0(p,q;T)\Biggr)\Biggr\}.
\end{eqnarray}

The details of the integration for the direct
coupling (5-b) including the full vector meson propagator is presented 
in  appendix C. The resulting contribution to  
the in medium pion electromagnetic form factor is
\begin{equation}
F^{direct}_\pi(T)=  {5T^2\over 4f_\pi^2}g_0(m_\pi^2/T^2)
          -3g^2\biggl[A(p;T)+f_0(p;T)\biggl],
\end{equation}
where $A(p;T)$ and $f_0(p;T)$ are given in appendix C.
We should note, however, that the correction to the result obtained in the
soft pion approximation is small for this contribution.

The biggest changes from the soft pion result   
arise from the pion wave function renormalization constant. 
In the soft pion limit we can see that there is a cancelation between the
contribution from $\pi-\rho$ meson loop diagram and a-dependent term in 
$\pi-$tadpole diagram. 
The pion wave function correction in hot matter is given by 
\begin{equation}
Z_\pi^{-1}=1-{2T^2\over3f_\pi^2}g_0(m_\pi^2/T^2)
\end{equation}
Thus, in the soft pion limit the pion 
tadpole diagram increases the wave function
renormalization constant. 

\begin{figure}
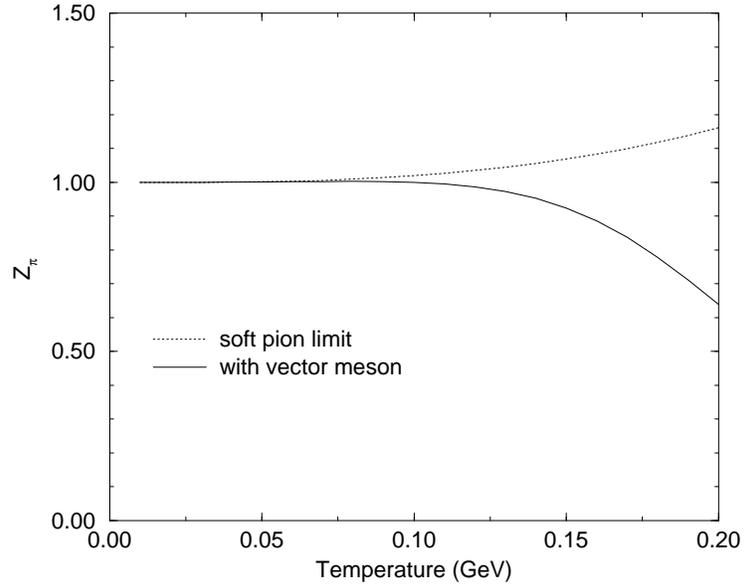

\caption{Pion wave function renormalization constant at finite temperature.}
\label{wave}
\end{figure}

When we include the vector meson propagator, the pion self-energy is given by 
\begin{eqnarray}
\Pi(p_0, p\to 0) &=& {c_1\over \pi^2}
        \int{\vert\vec l\vert^2d\vert\vec l\vert\over\omega(e^{\omega/T}-1)}
\cr
&&+{c_2\over \pi^2} \int{\vert\vec l\vert^2d\vert\vec l\vert\over\omega
(e^{\omega/T}-1)}
{1\over (p_0^2-m_\rho^2+m_\pi^2)^2
-4\omega^2 p_0^2}
\end{eqnarray}
where the coefficients $c_1$, and $c_2$ are given by 
\begin{eqnarray}
c_1&=&g^2-{1\over 12f_\pi^2}(5m_\pi^2+2(p_0^2+m_\pi^2))\cr 
c_2&=&-g^2(p_0^2-m_\rho^2+m_\pi^2)
          (2p_0^2-(m_\rho^2-2m_\pi^2)-(p_0^2-m_\pi^2)^2/m_\rho^2)
\end{eqnarray}
With the definition in (\ref{zpi}) we have 
\begin{eqnarray}
Z_\pi^{-1} &=& 1 +{1\over 6f_\pi^2}{1\over\pi^2}
        \int{\vert\vec l\vert^2d\vert\vec l\vert\over\omega(e^{\omega/T}-1)}
\cr
           & & -{g^2\over\pi^2} (3m_\rho^2-8m_\pi^2)
        \int{\vert\vec l\vert^2d\vert\vec l\vert\over\omega(e^{\omega/T}-1)}
                {1\over (m_\rho^2-2m_\pi^2)^2-4\omega^2 m_\pi^2}\cr
           & & +{g^2\over\pi^2} (m_\rho^2-2m_\pi^2)(m_\rho^2-4m_\pi^2)
        \int{\vert\vec l\vert^2d\vert\vec l\vert\over\omega(e^{\omega/T}-1)}
                {4\omega^2+2(m_\rho^2-2m_\pi^2)\over 
                 [(m_\rho^2-2m_\pi^2)^2-4\omega^2 m_\pi^2]^2}
\end{eqnarray}
The resulting $Z_\pi$ is shown in fig.~\ref{wave} together 
with the result obtained
in soft pion limit. We have an identical result at low temperatures, $T<100$
MeV in both cases.  However, as temperature increases the effect from 
vector mesons becomes important and the $Z_\pi$ begins to drop. This reduction
of $Z_\pi$, of course,  is simply due to the fact 
that parts of the pion wave function may now
reside `rho - thermal pion' states with the same quantum numbers as the pion.

\begin{figure}
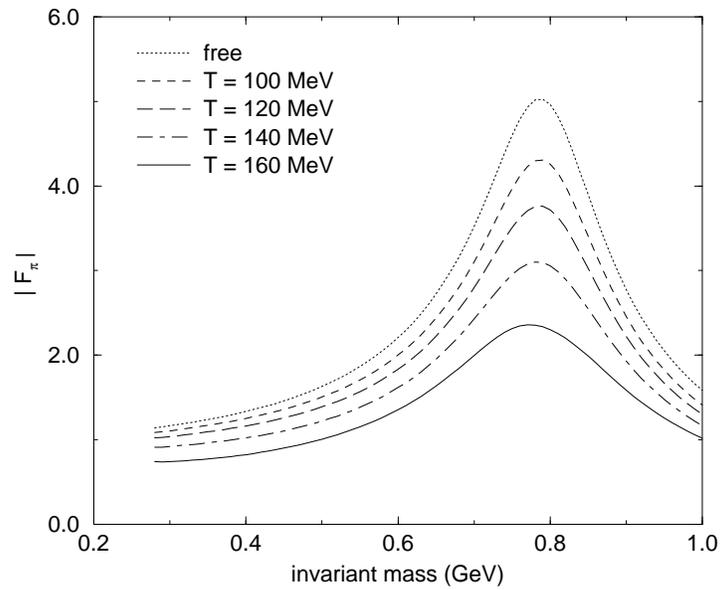

\caption{Pion electromagnetic form factor at finite temperature}
\label{form}
\end{figure}

Fig. \ref{form} shows the pion form factor around the
$\rho$ resonance for different temperatures.
The pion electromagnetic form factor is
seen to be further reduced near the resonance as the temperature increases.
We obtain a reduction of the form factor by 50\% at the invariant mass of
virtual photon $M\sim m_\rho$
when $T=160$ MeV. This result is comparable with that obtained 
using the QCD sum-rule approach which shows that 
at $M^2\sim (1 {\rm GeV})^2$  
the form factor at $T\sim$ 0.9 $T_c$ is about half its value at
$T=0$ \cite{domi}. 
It is also consistent with that based on the perturbative QCD at high $M^2$ 
\cite{satz}.

The reduction of the pion electromagnetic form factor 
at finite temperature
is related to chiral symmetry restoration and 
the deconfinement phase transition in hot hadronic matter.
The photon-$\rho$-meson coupling is modified due to the vector axial-vector 
mixing at finite temperature which has been regarded as a possible
signature for the 
partial restoration of chiral symmetry in hot matter \cite{ks}.
The resonance width has also been expected to increase in hot 
hadronic matter as the system undergoes chiral symmetry restoration 
and the deconfinement phase transition.
The vertex corrections, which lead to a reduction of the
rho-pion coupling constant at finite temperature, 
may be related to the recent suggestion that the
pion-vector meson coupling constant vanishes when chiral symmetry 
is restored in the vector limit \cite{rho}. 
The possible relation between the suppression of the form factor 
and phase transition in hot hadronic matter also has been suggested via 
QCD sum rules \cite{domi} and the QCD factorization formula \cite{satz}
to lead to similar suppressions in the pion electromagnetic form factor.

We should note that the form factor is not equal to 1 as the invariant mass 
approaches to zero.  
This, however, does not contradict the charge conservation discussed in the
previous section. Since we are working in the back to back frame, the
three-momentum of the virtual photon is zero. Therefore, going to the invariant
mass zero limit corresponds to the limit $k_0 \rightarrow 0, \,\, \vec{k} =
0$.
In this limit, the conserved charge derived from the WT identities is related
to $\Gamma_0$. The form factor, however, 
is proportional to the space component of the vertex function, $\Gamma_i$, 
since we are working in the frame where $p_0=q_0$ and $\vec p=-\vec q$.
A  similar behavior of the form factor also has been observed in
dense matter \cite{friman}.

\section{Dilepton emission from pion-pion annihilation}

In this section we consider the effect of the
medium corrections, 
especially the in-medium pion electromagnetic form factor, on the dilepton 
production from hot hadronic matter. This has been of  interest
because of recent dilepton measurements  
at SPS-energy heavy ion collisions.  Experiments measured by
the CERES collaboration at the CERN/SPS show a significant enhancement of
dileptons over a  hadronic cocktail  
in the invariant mass region  200 MeV $< M <$ 1500 MeV
in the S+Au collision at 200 GeV/nucleon \cite{ceres}.
On the other hand, in proton induced reactions 
such as the 450 GeV p-Be and p-Au
collisions, the low-mass dilepton spectra can be satisfactorily explained by
dileptons from hadron decays. 
The enhancement seen in the CERES experiment is for dileptons at central
rapidity where the charge particle density is high. In another experiment
by the HELIOS-3 collaboration \cite{helios}, dileptons at forward rapidities
were measured, where the charge particle density is low,  and the enhancement
was found to be smaller. Suggestions have thus been made that the excess
dileptons seen in these experiments are from
pion-pion annihilation, $\pi^+\pi^-\to e^+e^-$.
However, model calculations which have taken this channel into account can at
best reach the lower end of the sum of statistical and systematic errors of the
CERES-data in the low invariant mass region. For the HELIOS data, which
unfortunately do not show a systematic error, there is still a disagreement by
a factor of $\sim 1.5$ around an invariant mass of $500 \, \rm MeV$
\cite{gqli,cassing,volker}.

It is, therefore, interesting to see to which extent the in medium correction
modify the dilepton production. Here we will concentrate on the 
pion annihilation channel.  
With modified pion properties in medium 
the production rate of dileptons with vanishing three momentum
can be written as \cite{gale}
\begin{equation}
{d^4R\over d^3kdM}\Biggr\vert_{\vec{k}=0}={\alpha^2\over3(2\pi)^4}
{\vert F_\pi(M,T)\vert^2\over(e^{\omega/T}-1)^2}
\sum_{\vert\vec p\vert}{\vert\vec p\vert^4\over\omega^4}
\Biggl\vert{d\omega\over d\vert\vec p\vert}\Biggr\vert^{-1},
\label{dilep}
\end{equation}
where $M$ is the dilepton invariant mass. 
The momentum and energy of the pion are denoted by $\vert\vec p\vert$ 
and $\omega$,
respectively, and are related by its dispersion relation in the medium.
The last factor takes into account this effect.
The sum over $\vert\vec p\vert$'s is restricted by 
$\omega(\vert\vec p\vert)=M/2$.
We also include the in medium form factor obtained in the previous section. 

The pion dispersion relation at finite temperature is determined from the
equation  
\begin{equation}
p_0^2-\vec p^2-m_\pi^2-\Pi_\pi(p_0,\vec p)=0,
\end{equation}
where $\Pi_\pi(p_0,\vec p)$ is given in (\ref{pion_self}).
Since the pion self-energy also depends on the momentum and energy of pion,
the above equation should be solved self-consistently. 
The real part of the equation determines the dispersion relation of the pion in
medium, while the
imaginary part is related to the absorptive properties of pions in hot matter.
We have shown that the pion mass, which is defined as the pole position of the
propagator, decreases with temperature and the dispersion curve is softened
in the low momentum region at finite temperature \cite{song4,song5}.

\begin{figure}
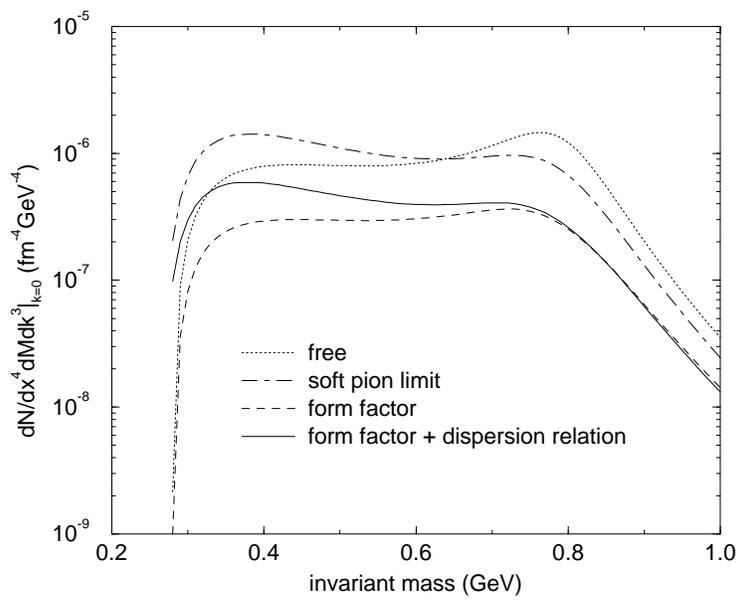

\caption{Dileptons from two pion annihilation at finite temperature}
\label{dilepton}
\end{figure}

With these medium effects on the pion dispersion relation and the form factor 
we get the dilepton production rate
as shown in Fig.~\ref{dilepton} for $T=$ 160 MeV.
The result obtained with the modified pion form factor  (dashed line) is 
compared with that calculated using the form factor in free space (dotted
line). Since the production rate is proportional to the  
square of the form factor, we obtain a larger reduction 
with temperature in the dilepton 
production rate than in the form factor above.
Near the $\rho$ meson resonance we have $dR[F_\pi(M,T)]$ $\sim$ 
$(1/2)^2$ $dR[F_\pi(M,0)]$ at $T=160$ MeV, 
and the dilepton production rate is 
reduced by almost a factor of four.
Finally, when we include the effect from the dispersion relation of pions we 
obtain the solid line. 
There are two prominent effects due to changes of the
pion dispersion relation in hot hadronic matter. 
First, the threshold of dilepton production
from pion-pion annihilation is lowered because of the reduction of the
pion mass, $m_\pi^*(T)$, at finite temperature.
Secondly, the dilepton production rate is
enhanced in the invariant mass region, $2m^*_\pi(T)<M<m_\rho$,
and shows a maximum at $M\sim 350$ MeV,
which is due to the softening of the pion dispersion relation in medium.
The latter, however, does not have any effect on dileptons with invariant
masses near the vector meson resonance.
For comparison we also show the result using the form factor obtained 
in the soft pion limit as dot-dashed line.

We see that there is a
significant suppression near the vector resonance region. 
Since the production rate around the two pion threshold, $M\sim 2m_\pi$ is not
changed due to the additional effect of the modified dispersion relation,
we have a relative enhancement at the
low invariant mass region, which is in qualitative agreement with the CERES
data. 
However, in order to compare with experiment, we need to
include properly the expansion dynamics of the hot matter that is formed in
high energy nucleus-nucleus collisions as well as the contribution from all
other channels and the experimental acceptance. 

In a recent work we have included these medium effects on 
dilepton production from hot hadronic matter in a hadronic transport model
\cite{volker}. In this calculation 
the result obtained from the calculation in the soft pion limit has been used 
in order to generate  the largest possible effect of the medium 
correction in the low invariant mass region. 
We found that in the total spectrum the in-medium effect is hardly 
visible, especially in the interesting low invariance mass region.
This is simply due to the fact that the pion
annihilation contributions less than 1/3 to the total yield in this region and
even an enhancement of a factor of two would increase the total spectrum by
less than 30\%. 

These medium effects, however, might be observable in the dilepton spectrum 
from 
the mixed phase. If there is a phase transition and
the hadronic system goes through a long lived mixed phase  
before freeze-out,   
the most important contribution to dilepton production 
would come from the hadronic component of the mixed phase 
at $T_c=160\sim 180$ MeV.
In this case our results imply a significant suppression in the production rate
of lepton pairs with invariant mass near the vector-meson mass. 
Moreover, when we assume that the  mixed phase expands very slowly \cite{sz} 
and, hence, produces more dileptons the effect is more significant and may
induce a large suppression due to the modification of the form factor.

\section{Summary}
 
In summary, we have studied the pion electromagnetic 
form factor in hot hadronic matter 
using an effective Lagrangian with vector mesons. 
In this model pions couple to the photon only through vector
mesons, according to vector meson dominance assumption.

We have considered leading corrections for the pion-photon coupling at
finite temperature. 
We have shown 
that in the soft-pion limit the Ward-Takahashi identity is satisfied.
While the WT identity implies charge conservation at zero
temperature, it is not straightforward in medium. We have considered two
different limits and define an effective charge separately. We could show that
this effective charge is conserved in each case.
 
We furthermore have  studied the pion electromagnetic 
form factor in time-like region at
non-zero temperature.  
We could show that there is a reduction in the magnitude of the form 
factor, which can be 
understood in terms of the partial restoration of chiral symmetry and 
the deconfinement transition in hot hadronic matter.  The reduction in the
electromagnetic form factor 
leads to a suppression of dilepton production from  
two pion annihilation in hot matter.  
However, this suppression is hardly visible in full spectrum because of the
contribution from other channels. 
Only when the $\pi-\pi$ contribution is dominant, for example in long-lived
mixed phase, one may be able to observe a 
medium effect on the pion electromagnetic form factor.

We expect various observable consequences of these medium effects on the
electromagnetic couplings of pions and vector mesons. For example, 
the ratio $N_\phi/(N_\rho+N_\omega)$ of the produced vector mesons, which is
extracted from dilepton measurements  would be modified \cite{slee,Nphi}.
Our results show that the number of dileptons from rho-meson decay
will be suppressed because of the reduced coupling to the photon in medium.
For  the $\omega$-meson, on the other hand,  there is no such effect. 
Also, thermal corrections to the phi decay should be small, 
since these involve kaons.
Consequently,  the ratio of dileptons from phi decay 
over those from omega and rho decay would show an enhancement even if the
actual particle ratios are unchanged. Therefore, when extracting the particle
ratios from the dilepton yields one should not conclude an enhancement of phi
mesons before the corrections to the rho-photon couplings have been properly
taken into account.
In this context it is of interest to extend present calculations to 
the $SU(3)$ limit and
to study the temperature dependence of the photon-phi meson coupling. 
It is also very interesting to study the form factor
needed in $\bar K-K$ annihilation. This will be relevant to
the double phi meson peak in the
dilepton spectrum, which has recently been suggested as a possible 
signal for the phase transition in hot matter \cite{yuki}. 
The second phi peak in the dilepton spectrum is from the decay of phi mesons
in the mixed phase, which have reduced masses as a result of 
partial restoration of chiral symmetry. 

CS would like to thank C. M. Ko and S. H. Lee for valuable conversation 
at the beginning of this work.
VK thanks  B. Friman for a useful discussion about the singular behavior
of the form factor in dense matter.  
This work supported by the Director, Office of Energy Research, Office of High
Energy and Nuclear Physics, Division of Nuclear Physics, Division of Nuclear
Sciences, of 
the U. S. Department of Energy under Contract No. DE-AC03-76SF00098.

\eject
\newpage

\begin{center} 
{\Large\bf Appendix A: Vertex correction $\Gamma_\mu^{(4-c)}$}
\end{center}

\noindent
When we assume that 
\begin{equation}
\Gamma_\mu\equiv (p_\mu-q_\mu)F,
\end{equation}
$F$ is given by
\begin{equation}
F={1\over (p-q)^2}(p^\mu-q^\mu)\Gamma_\mu.
\end{equation}
For the $\Gamma^{(4-c)}_\mu$ we have 
\begin{equation}
\Gamma_\mu^{(4-c)}=(p_\mu-q_\mu){1\over2} g^2 F_\pi(T=0)
                   {1\over (p-q)^2}\bar F^{(4-c)},
\end{equation}
where $F_\pi(T=0)$ is pion form factor in free space and 
\begin{eqnarray}
\bar F^{(4-c)} &=& T\sum_{n_l}\int{d^3l\over(2\pi)^3}\Biggl\{
 {l\cdot(p-q)\over(l^2-m_\pi^2)((l-p)^2-m_\pi^2)}
 \nonumber \\ [12pt] 
&&+{l\cdot(p-q)\over(l^2-m_\pi^2)((l+q)^2-m_\pi^2)}
 \nonumber \\ [12pt] 
&&+(2m_\pi^2-3k^2/2)
 {2l\cdot(p-q)\over(l^2-m_\pi^2)((l-k)^2-m_\pi^2)((l-p)^2-m_\rho^2)}
 \nonumber \\ [12pt] 
&&+{2(l\cdot(p-q))^2\over(l^2-m_\pi^2)((l-k)^2-m_\pi^2)((l-p)^2-m_\rho^2)}
\Biggr\}
 \nonumber \\ [12pt] 
&\equiv&f_1(p,q;T)+f'_1(p,q;T)+2(2m_\pi^2-3k^2/2)g_1(p,q;T)+2g_2(p,q;T).
\end{eqnarray}

The loop integration can be done in back-to-back frame where $p_0=q_0$ and
$\vec p=-\vec q$.

{\bf (i)} The function $f_1$ and $f'_1$ are given by 
\begin{eqnarray}
f_1(p,q;T) &=& T\sum_{n_l}\int{d^3l\over(2\pi)^3}
           {l\cdot (p-q)\over (l^2-m_\pi^2)((l-p)^2-m_\rho^2)}
\nonumber \\ [12pt] 
          &=& {1\over 8\pi^2}\int 
{\vert\vec l\vert^2d\vert\vec l\vert\over \omega(e^{\omega/T}-1)}
           \left(4-{A_{+}\over 2\vert\vec p\vert\vert\vec l\vert}L_{+}
                  -{A_{-}\over 2\vert\vec p\vert\vert\vec l\vert}L_{-}\right),
\nonumber \\ [12pt] 
f'_1(p,q;T) &=& T\sum_{n_l}\int{d^3l\over(2\pi)^3}
               {l\cdot (p-q)\over (l^2-m_\pi^2)((l+q)^2-m_\rho^2)}
\nonumber \\ [12pt] 
          &=& f_1(p,q;T),
\end{eqnarray}
where 
\begin{equation}
L_{\pm}=\ln \left[{A_\pm+2\vert\vec p\vert\vert\vec l\vert\over 
                   A_\pm-2\vert\vec p\vert\vert\vec l\vert}\right],
\end{equation}
with
\begin{equation}
A_\pm=(m_\rho^2-2m_\pi^2)\pm 2\omega p_0.
\end{equation}

{\bf (ii)} The functions $g_n$'s can also be obtained and are given by
\begin{eqnarray}
g_1(p,q;T) &=& T\sum_{n_l}\int{d^3l\over(2\pi)^3}
           {l\cdot (p-q)\over (l^2-m_\pi^2)((l-k)^2-m_\pi^2)((l-p)^2-m_\rho^2)}
\nonumber \\ [12pt]           
 &=& {1\over 4\pi^2}
\int{\vert\vec l\vert^2d\vert\vec l\vert\over\omega (e^{\omega/T}-1)}
\Biggl\{
{1\over 4p_0}\left({1\over p_0+\omega}+{1\over p_0-\omega}\right)
             \left(2-{A_{+}\over 2\vert\vec p\vert\vert\vec l\vert}L_{+}\right)
\nonumber \\ [12pt] 
&&\qquad\qquad\qquad
+{1\over 4p_0}{1\over p_0+\omega}
              \left({A_{+}\over 2\vert\vec p\vert\vert\vec l\vert}L_{+}
                   -{A_{-}\over 2\vert\vec p\vert\vert\vec l\vert}L_{-}\right)
\Biggr\},
\end{eqnarray}
and 
\begin{eqnarray}
g_2(p,q;T) &=& T\sum_{n_l}\int{d^3l\over(2\pi)^3}
     {(l\cdot(p-q))^2\over (l^2-m_\pi^2)((l-k)^2-m_\pi^2)((l-p)^2-m_\rho^2)}
\nonumber \\ [12pt] 
          &=& {1\over 4\pi^2}
\int{\vert\vec l\vert^2d\vert\vec l\vert\over\omega (e^{\omega/T}-1)}
\Biggl\{
{A_{+}\over 4p_0}\left({1\over p_0+\omega}+{1\over p_0-\omega}\right)
                 \left(2-{A_{+}\over 2\vert\vec p\vert\vert\vec l\vert}L_{+}
\right)
\nonumber \\ [12pt] 
&&\qquad
-{1\over 4p_0}{1\over p_0+\omega}
\left[ A_{+}\left(2-{A_{+}\over 2\vert\vec p\vert\vert\vec l\vert}L_{+}\right)
      -A_{-}\left(2-{A_{-}\over 2\vert\vec p\vert\vert\vec l\vert}L_{-}\right)
\right]
\Biggr\}.
\end{eqnarray}

\bigskip
\begin{center} 
{\Large\bf Appendix B: Vertex correction $\Gamma_\mu^{(4-d)}$}
\end{center}

\noindent
With the same approximation we can also calculate the contribution from the
diagram (4-d) in the back-to-back frame;
\begin{equation}
\Gamma^{(4-d)}=(p_\mu-q_\mu)2g^2 F_\pi(T=0) \bar F^{(4-d)},
\end{equation}
where
\begin{eqnarray}
\bar F^{(4-d)} &=& -T\sum_{n_l}\int{d^3l\over(2\pi)^3}\Biggl\{
{1\over(p-q)^2} {2(l\cdot(p-q))^2-(k^2+4p\cdot q) l\cdot(p-q)
\over(l^2-m_\pi^2)((l-p)^2-m_\rho^2)((l+q)^2-m_\rho^2)}
 \nonumber \\ [12pt] 
&&\qquad\qquad
+{l\cdot(p-q)\over(l^2-m_\pi^2)((l-p)^2-m_\rho^2)((l+q)^2-m_\rho^2)}
 \nonumber \\ [12pt] 
&&\qquad\qquad+(2m_\pi^2-3k^2/2)
 {1\over(l^2-m_\pi^2)((l-p)^2-m_\rho^2)((l+q)^2-m_\rho^2)}
\Biggr\}
 \nonumber \\ [12pt] 
&\equiv& {2\over(p-q)^2}h_2(p,q;T)
        +\left[1-{k^2+4p\cdot q\over(p-q)^2}\right] h_1(p,q;T)
        +(2m_\pi^2-3k^2/2)h_0(p,q;T)\nonumber\\.
\end{eqnarray}
The function $h_n$'s are defined as 
\begin{equation}
h_n=T\sum_{n_l}\int{d^3l\over(2\pi)^3}
{(l\cdot(p-q))^n\over(l^2-m_\pi^2)((l-p)^2-m_\rho^2)((l+q)^2-m_\rho^2)}.
\end{equation}
Each integration has been done and is given by
\begin{eqnarray}
h_0 &=& {1\over32\pi^2p_0\vert\vec p\vert}
    \int{\vert\vec l\vert d\vert\vec l\vert\over\omega^2(e^{\omega/T}-1)}
    (L_{+}-L_{-}),
\nonumber \\ [12pt] 
h_1 &=& -f_0(p;T)-(m_\rho^2-2m_\pi^2)h_0(p,q;T),
\nonumber \\ [12pt] 
h_2 &=& (m_\rho^2-2m_\pi^2)[f_0(p;T)+(m_\rho^2-2m_\pi^2)h_0(p,q;T)]-f_1(p,q;T),
\end{eqnarray}
where $L_{\pm}$ and $f_1$ are given in appendix A
and $f_0(p;T)$ is defined by 
\begin{eqnarray}
f_0(p;T) &=& T\sum_{n_l}\int{d^3l\over(2\pi)^3}
           {1\over (l^2-m_\pi^2)((l-p)^2-m_\rho^2)}  
\nonumber \\ [12pt] 
     & =& {1\over 8\pi^2}
\int {\vert\vec l\vert^2d\vert\vec l\vert\over \omega(e^{\omega/T}-1)}
       {1\over 2\vert\vec p\vert\vert\vec l\vert}(L_{+}+L_{-}).
\end{eqnarray}

\bigskip
\begin{center} 
{\Large\bf Appendix C: Direct coupling in medium $\Gamma^{(5-b)}_\mu$}
\end{center}

\noindent
For the direct coupling of pions to photon fields in medium we have 
\begin{equation}
\Gamma^{(5-b)}_\mu=-3g^2 T\sum_{n_l}\int{d^3l\over(2\pi)^3}
\Biggl[ {(l^\mu+p^\mu)\over (l^2-m_\pi^2)((l-p)^2-m_\rho^2)}
\Biggr].
\end{equation}
It can be written as 
\begin{equation}
\Gamma^{(5-b)}_\mu=-3g^2
\biggl[g_\mu(p;T)+p_\mu f_0(p;T)\biggr],
\end{equation}
where $f_0(p;T)$ is given in Appendix B and 
\begin{equation}
g_\mu(p;T) = T\sum_{n_l}\int{d^3l\over(2\pi)^3}
           {l_\mu\over (l^2-m_\pi^2)((l-p)^2-m_\rho^2)}.
\end{equation}

For $g_\mu(p;T)$ it can be written in general as
\begin{equation}
g_\mu(p;T)=A(p;T) p_\mu+B(p;T)p_0 n_\mu
\end{equation}
with
\begin{eqnarray}
A &=& -{1\over \vert\vec p\vert^2}(p^\mu g_\mu-p^0 g_0),
\cr
B &=& g_0/p_0-A.
\end{eqnarray}
In back-to-back frame we have 
\begin{eqnarray}
A(p;T) &=& -{1\over 16\pi^2\vert\vec p\vert^2}
\int{\vert\vec l\vert^2d\vert\vec l\vert\over\omega(e^{\omega/T}-1)}
\left[4
-{(m_\rho^2-2m_\pi^2)\over 2\vert\vec p\vert\vert\vec l\vert}(L_{+}+L_{-})
-{p_0\omega\over \vert\vec p\vert\vert\vec l\vert}(L_{+}-L_{-})\right],
\nonumber \\ [12pt] 
B(p;T) &=& {1\over 8\pi^2 p_0}
\int{\vert\vec l\vert^2d\vert\vec l\vert\over(e^{\omega/T}-1)}
{1\over 2\vert\vec p\vert\vert\vec l\vert}(L_{+}-L_{-})-A. 
\end{eqnarray}

By the same way we have 
\begin{equation}
\Gamma^{(5-c)}_\mu=3g^2
\biggl[g_\mu(q;T)+q_\mu f_0(q;T)\biggr],
\end{equation}
with 
\begin{equation}
g_\mu(q;T)=A(q;T) q_\mu+B(q;T)q_0 n_\mu
\end{equation}
Since $p_0=q_0$ and $\vert\vec p\vert=\vert\vec q\vert$ in the frame we are
working, $f_0(p;T)=f_0(q;T), A(p;T)=A(q;T)$ and $B(p;T)=B(q;T)$.
Thus 
\begin{equation}
\Gamma^{(5-b)}_\mu+\Gamma^{(5-c)}_\mu 
= -3g^2 (p_\mu-q_\mu)(A(p;T)+f_0(p;T)).
\end{equation}
\eject

\newpage



\begin{thebibliography}{99}


\bibitem{qgp} E. V. Shuryak, Phys. Report  {\bf 61}, 71 (1980).

\bibitem{rhic} For recent review, see {\it Proc. Quark Matter '95}, 
               Nucl. Phys. {\bf A590}, 103c (1995).

\bibitem{dilepton} E. L. Feinberg, Nuovo Cimento {\bf 34A}, 39 (1976).\\
                   E. V. Shuryak, Sov. J. Nucl. Phys. {\bf 28}, 408 (1978).
 
\bibitem{pisaski} R. Pisarski, Phys. Lett. B {\bf 110}, 155 (1982).

\bibitem{br}  G. Brown and M. Rho, Phys. Rev. Lett. {\bf 66}, 2720 (1991).  

\bibitem{dei} M. Dey, V. L. Eletsky and B. L. Ioffe, 
              Phys. Lett. B {\bf 252}, 620 (1990).

\bibitem{slee} S. H. Lee, C. Song and H. Yabu,
                Phys. Lett. {\bf B341}, 407 (1995).

\bibitem{song1} Chungsik Song, Phys. Rev. D {\bf 53}, 3962 (1996).

\bibitem{pisaski2} R. Pisarski, Phys. Rev. D {\bf 52}, 4694 (1994).

\bibitem{song2} Chungsik Song, S. H. Lee and C. M. Ko,
                Phys. Rev. C {\bf 52}, R476 (1995).

\bibitem{sak} J. J. Sakurai, {\it Currents and Mesons} 
              (Chicago, Chicago, 1969).

\bibitem{hls} M. Bando, T. Kugo and K. Yamawaki, 
              Phys. Rep. {\bf 164}, 217 (1988).

\bibitem{ks} J. Kapusta and E. Shuryak, Phys. Rev. D {\bf 49}, 4694 (1994).  

\bibitem{drell} J. Bjorken and S. Drell, {\it Relativistic Quantum Mechanics}
                (McGraw-Hill, New York, 1964).

\bibitem{song3} Chungsik Song, Phys. Rev. D {\bf 48}, 1375 (1993).


\bibitem{kevin} K. Haglin, Nucl. Phys. A {\bf 584}, 719 (1995).

\bibitem{kk} K. Kajantie and J. Kapusta,
             Ann. Phys. (N.Y.) {\bf 160}, 477 (1985).

\bibitem{heinz} U. Heinz, K. Kajantie and T. Toimela, 
                Ann. Phys. (N.Y.) {\bf 176}, 218 (1987).

\bibitem{domi} C. Dominguez, M. Loewe and J. Rozowsky, 
               Phys. Lett. B {\bf 335}, 506 (1994).
 
\bibitem{satz} D. Kharzeev and H. Satz, Phys. Lett B {\bf 340}, 167 (1994).
  
\bibitem{friman} M. Herrmann, B. L. Friman and W. N\"orenberg,
                 Nucl. Phy. B {\bf 560}, 411 (1993).

\bibitem{rho}  G. Brown and M. Rho, Phys. Lett B {\bf 338}, 301 (1994).  
 
\bibitem{ceres} G. Agakichiev et al., Phys. Rev. Lett. {\bf 75} 1272 (1995);\\
                J.P. Wurm for the CERES/NA45 collaboration,
                Nucl. Phys. {\bf A590 } 103c (1995);\\
                A. Drees for the CERES/NA45 collaboration,
                {\em Proc. International Workshop XXIII on Gross Properties
                 of Nuclei and Nuclear Excitations}, 
                ed. H. Feldmeier and W. N\"orenberg, 
                (GSI, Darmstadt 1995) p. 151.

\bibitem{helios} M. Masera for the HELIOS-3 Collaboration, 
                 in {\it Proc. Quark Matter '95}, January 9-13, 1995; 
                 Nucl. Phys. A, to be published.

\bibitem{gqli} G. Li, C. M. Ko and G. E. Brown, 
               Phys. Rev. Lett. {\bf 75}, 4007 (1995).\\
               D. K. Srivastava, B. Sinha and C. Gale, 
               McGill University Preprint (1995). 

\bibitem{cassing} W.Cassing, W. Ehehalt, and C.M. Ko, Phys. Lett. B {\bf 363},
                  379 (1996)\\
                  W. Cassing, W. Ehehalt, and I. Kralic, 
                  Phys. Lett. B {\bf 377}, 5 (1996). 

\bibitem{volker} Volker Koch and Chungsik Song, Report No. LBL-38619 (1996),
                 submitted to Phys. Rev. C.

\bibitem{gale}  C. Gale and J. Kapusta, Phys. Rev. C {\bf 38}, 2659 (1988).
 
\bibitem{song4} Chungsik Song, Phys. Rev. D {\bf 49}, 1556 (1994).

\bibitem{song5} C. Song, V. Koch, S. H. Lee and C. M. Ko,
                Phys. Lett. B {\bf 366}, 379 (1996).

\bibitem{sz} C. M. Hung and E. Shuryak, Phys. Rev. Lett. 75, 4003 (1995).

\bibitem{Nphi} NA38 collaboration, C. Baglin et al., 
               Phys. Lett. B 272, 449 (1991)

\bibitem{yuki} H. Asakawa and C. M. Ko, Phys. Lett. B {\bf 322}, 33 (1994).

\end{thebibliography}
\end{document}